\RequirePackage{lineno}
\documentclass[aps,prc,twocolumn,showpacs,superscriptaddress,groupedaddress, ]{revtex4}  
\usepackage{graphicx}  
\usepackage{amssymb}
\usepackage{subfigure}

\begin{document}

\title{Influence of $\phi$ mesons on negative kaons in Ni+Ni collisions at 1.91A GeV beam energy}

\def\wars{Institute of Experimental Physics, Faculty of Physics, University of Warsaw, Warsaw, Poland}
\def\heid{Physikalisches Institut der Universit\"{a}t Heidelberg, Heidelberg, Germany}
\def\darm{GSI Helmholtzzentrum f\"{u}r Schwerionenforschung GmbH, Darmstadt, Germany}
\def\seou{Korea University, Seoul, Korea}
\def\cler{Laboratoire de Physique Corpusculaire, IN2P3/CNRS, and Universit\'{e} Blaise Pascal, Clermont-Ferrand, France}
\def\zagr{Ru{d\llap{\raise 1.22ex\hbox{\vrule height 0.09ex width 0.2em}}\rlap{\raise 1.22ex\hbox{\vrule height 0.09ex width 0.06em}}}er Bo\v{s}kovi\'{c} Institute, Zagreb, Croatia}
\def\munI{Excellence Cluster Universe, Technische Universit\"{a}t M\"{u}nchen, Garching, Germany}
\def\munII{E12, Physik Department, Technische Universit\"{a}t M\"{u}nchen, Garching, Germany}
\def\vien{Stefan-Meyer-Institut f\"{u}r subatomare Physik, \"{O}sterreichische Akademie der Wissenschaften, Wien, Austria}
\def\sp{University of Split, Split, Croatia}
\def\buda{Wigner RCP, RMKI, Budapest, Hungary}
\def\mosc{Institute for Theoretical and Experimental Physics, Moscow, Russia}
\def\dres{Institut f\"{u}r Strahlenphysik, Helmholtz-Zentrum Dresden-Rossendorf, Dresden, Germany} 
\def\harb{Harbin Institute of Technology, Harbin, China}
\def\kurc{Kurchatov Institute, Moscow, Russia}
\def\buch{Institute for Nuclear Physics and Engineering, Bucharest, Romania}
\def\stra{Institut Pluridisciplinaire Hubert Curien and Universit\'{e} de Strasbourg, Strasbourg, France}
\def\tsing{Department of Physics, Tsinghua University, Beijing 100084, China}
\def\lan{Institute of Modern Physics, Chinese Academy of Sciences, Lanzhou, China}

\author{K.~Piasecki}\email{krzysztof.piasecki@fuw.edu.pl} \affiliation{\wars}
\author{N.~Herrmann} \affiliation{\heid}
\author{R.~Averbeck} \affiliation{\darm}
\author{A.~Andronic} \affiliation{\darm} 
\author{V.~Barret} \affiliation{\cler} 
\author{Z.~Basrak} \affiliation{\zagr} 
\author{N.~Bastid} \affiliation{\cler}
\author{M.L.~Benabderrahmane} \affiliation{\heid}
\author{M.~Berger} \affiliation{\munI} \affiliation{\munII}
\author{P.~Buehler} \affiliation{\vien} 
\author{M.~Cargnelli} \affiliation{\vien} 
\author{R.~\v{C}aplar} \affiliation{\zagr}
\author{P.~Crochet} \affiliation{\cler} 
\author{O.~Czerwiakowa} \affiliation{\wars}
\author{I.~Deppner} \affiliation{\heid}
\author{P.~Dupieux} \affiliation{\cler}
\author{M.~D\v{z}elalija} \affiliation{\sp}
\author{L.~Fabbietti} \affiliation{\munI} \affiliation{\munII}
\author{Z.~Fodor} \affiliation{\buda}
\author{P.~Gasik} \affiliation{\wars}
\author{I.~Ga\v{s}pari\'c} \affiliation{\zagr}
\author{Y.~Grishkin} \affiliation{\mosc}
\author{O.N.~Hartmann} \affiliation{\darm}
\author{K.D.~Hildenbrand} \affiliation{\darm}
\author{B.~Hong} \affiliation{\seou}
\author{T.I.~Kang}\affiliation{\darm}\affiliation{\seou}
\author{J.~Kecskemeti} \affiliation{\buda}
\author{Y.J.~Kim} \affiliation{\darm}
\author{M.~Kirejczyk} \affiliation{\wars}
\author{M.~Ki\v{s}} \affiliation{\darm} \affiliation{\zagr}
\author{P.~Koczon} \affiliation{\darm}
\author{R.~Kotte} \affiliation{\dres}
\author{A.~Lebedev} \affiliation{\mosc}
\author{Y.~Leifels} \affiliation{\darm}
\author{A.~Le F\`{e}vre} \affiliation{\darm}
\author{J.L.~Liu} \affiliation{\heid} \affiliation{\harb}
\author{X.~Lopez} \affiliation{\cler}
\author{V.~Manko} \affiliation{\kurc}
\author{J.~Marton} \affiliation{\vien}
\author{T.~Matulewicz} \affiliation{\wars}
\author{R.~M\"{u}nzer} \affiliation{\munI} \affiliation{\munII}
\author{M.~Petrovici} \affiliation{\buch}
\author{F.~Rami} \affiliation{\stra}
\author{A.~Reischl} \affiliation{\heid}
\author{W.~Reisdorf} \affiliation{\darm}
\author{M.S.~Ryu} \affiliation{\seou}
\author{P.~Schmidt} \affiliation{\vien}
\author{A.~Sch\"{u}ttauf} \affiliation{\darm}
\author{Z.~Seres} \affiliation{\buda}
\author{B.~Sikora} \affiliation{\wars}
\author{K.S.~Sim} \affiliation{\seou}
\author{V.~Simion} \affiliation{\buch}
\author{K.~Siwek-Wilczy\'{n}ska} \affiliation{\wars}
\author{V.~Smolyankin} \affiliation{\mosc}
\author{K.~Suzuki} \affiliation{\vien}
\author{Z.~Tymi\'{n}ski} \affiliation{\wars}
\author{P.~Wagner} \affiliation{\stra}
\author{I.~Weber} \affiliation{\sp} 
\author{E.~Widmann} \affiliation{\vien}
\author{K.~Wi\'{s}niewski} \affiliation{\heid} \affiliation{\wars}
\author{Z.G.~Xiao} \affiliation{\tsing}
\author{I.~Yushmanov} \affiliation{\kurc}
\author{Y.~Zhang} \affiliation{\heid} \affiliation{\lan}
\author{A.~Zhilin} \affiliation{\mosc}
\author{V.~Zinyuk} \affiliation{\heid}
\author{J.~Zmeskal} \affiliation{\vien}

\collaboration{FOPI Collaboration} \noaffiliation

\date{\today}

\begin{abstract}
$\phi$ and K$^-$ mesons from Ni+Ni collisions at the beam energy of 1.91A GeV
have been measured by the FOPI spectrometer, with a trigger selecting central 
and semi-central events amounting to 51\% of the total cross section. The 
phase space distributions, and the total yield of K$^-$, as well as the kinetic 
energy distribution and the total yield of $\phi$ mesons are presented. The 
$\phi$\slash K$^-$ ratio is found to be 
$0.44 \pm 0.07(\text{stat}) ^{+0.18}_{-0.12} (\text{syst})$, meaning that 
about 22\% of K$^-$ mesons originate from the decays of $\phi$ mesons, 
occurring mostly in vacuum. The inverse slopes of direct kaons are up to 
about 15 MeV larger than the ones extracted within the one-source model, 
signalling that a considerable share of gap between the slopes of K$^+$ and 
K$^-$ could be explained by the contribution of $\phi$ mesons to negative 
kaons. 
\end{abstract}

\maketitle

\section{Introduction}

Nucleus-nucleus collisions at the beam kinetic energies of 1-2A GeV offer the
unique possibility to study the onset of the strangeness production. 
The emergence of strangeness at beam energies below the thresholds in free 
nucleon-nucleon (NN) collisions is facilitated by the appearance of resonances
and mesons in the heated (T $\approx 100$~MeV) and compressed (2-3 times above 
the normal nuclear density $\rho_0$) collision zone, where the basic 
properties of particles like effective mass and decay constant are 
modified~\cite{Fuch06,Lutz04,Hart11,Hong97,Wisn00,Fors07,Lop07}.

The knowledge about the emission yields and kinematical properties of $\phi$ 
mesons produced in nucleus-nucleus collisions is very limited at the beam 
energy below 10A GeV as the experimental data is scarce in this beam energy 
region~\cite{Mang02,Agak09,Gasi10,Back04}. Possible channels of in-medium 
$\phi$ meson production include BB~$\rightarrow$~BB$\phi$, 
MB~$\rightarrow$~N$\phi$, $\rho\phi \rightarrow \phi$ (B = [N,~$\Delta$], 
M = [$\rho$,~$\pi$]). 

Calculations within the BUU transport model for the central Ni+Ni collisions 
at the beam energy of 1.93A GeV favour the dominance of the MB 
channels~\cite{Barz02}. This system was measured in a previous work by the 
FOPI Collaboration, but the statistics (23$\pm$7 events attributed to 
$phi$ mesons) was insufficient for a quantitative comparison with the 
calculations~\cite{Mang02}. 

\begin{figure*}
 \begin{minipage}{0.495\textwidth}
  \begin{center}
   \includegraphics[width=0.95\textwidth]{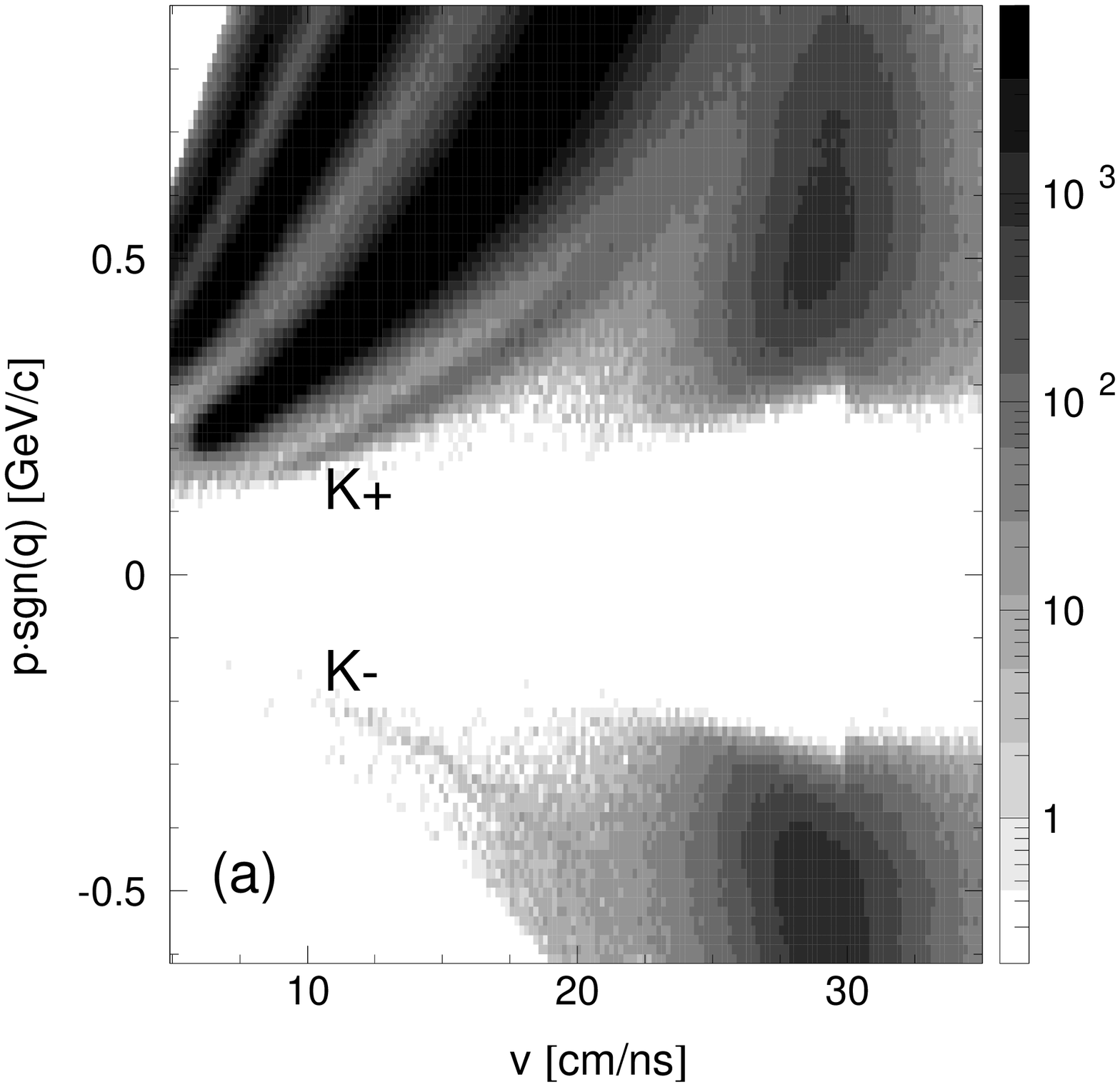}
  \end{center}
 \end{minipage}
 \begin{minipage}{0.495\textwidth}
  \begin{center}
   \includegraphics[width=0.95\textwidth]{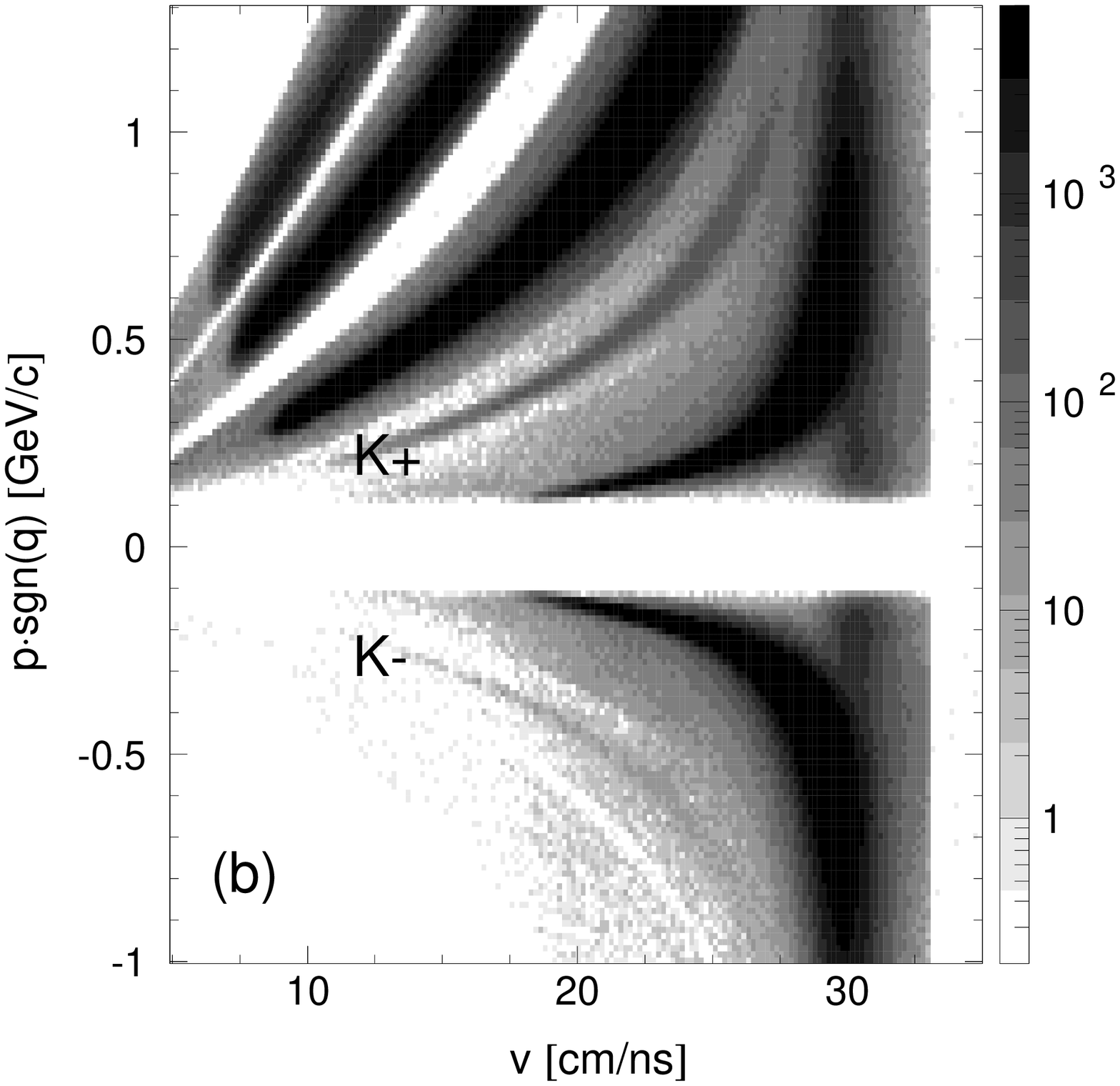}
  \end{center}
 \end{minipage}
 \caption{\label{fig:pvplot}Particle identification based on relativistic 
          momentum-velocity dependence, with (a) PSB, and (b) MMRPC detectors}
\end{figure*}

Since the mean decay path of $\phi$ is 46~fm, most of these particles decay 
outside the collision zone. As its dominant decay channel is 
$\phi \rightarrow K^+K^-$ (BR = 48.9\%)~\cite{PDG}, and the freeze-out yields 
of $\phi$ and K$^-$ are found to be of comparable 
order~\cite{Mang02,Agak09,Gasi10}, $\phi$ decays are the source of K$^-$ mesons
that are mostly unaffected by presence inside the medium, in contrast to those 
negative kaons produced directly in the collision zone. Therefore, evaluation 
of the $\phi$\slash K$^-$ ratio is of importance for the studies of the 
modifications of K$^-$ properties in-medium. In the discussed energy range 
this ratio was reported for the Ar+KCl collisions at 1.756A GeV~\cite{Agak09}. 

In this paper we present the yield and kinetic energy distribution of $\phi$, 
as well as the $\phi$\slash K$^-$ ratio for the collisions of Ni+Ni at the 
beam kinetic energy of 1.91A GeV, covering the most central 51\% of the 
geometrical cross section.

\section{The experiment}

The experiment was carried out with the FOPI spectrometer, installed at the 
heavy-ion synchrotron SIS-18 in GSI, Darmstadt. The innermost detector is the 
azimuthally symmetric Central Drift Chamber (CDC), covering the wide range 
of polar angles ($27^\circ < \theta_{\text{lab}} < 113^\circ$). CDC is 
surrounded by two detectors in the barrel geometry, dedicated for the 
Time-of-Flight (ToF) measurements: Multi-strip Multi-gap Resistive Plate 
Counter~\cite{Kis11}, spanning $30^\circ < \theta_{\text{lab}} < 53^\circ$, 
and the Plastic Scintillation Barrel (PSB), covering 
$55^\circ < \theta_{\text{lab}} < 110^\circ$. These devices are encircled by 
the magnet solenoid, delivering the magnetic field of B = 0.617~T, and 
covered at front by the Plastic scintillation Wall (PlaWa). More details on 
characteristics and performance of the FOPI apparatus can be found 
in~\cite{FOPI}. 

The $^{58}$Ni ions, accelerated to the kinetic energy of 1.91A GeV, were 
incident on the 405~$\mu$m-thick $^{58}$Ni target (corresponding to 1\% 
interaction probability). By requiring the multiplicities of charged hits in 
PlaWa (PSB) to be $\geq$~5 ($\geq$~1), the trigger selected the sample of 
$7.6\times 10^7$ central and semi-central events amounting to 51\% of the 
total geometrical cross section. Assuming the simple geometrical model of 
interpenetrating spheres, and the sharp cut-off approximation between the 
maximum impact parameter and the total reaction cross section, the mean 
number of participant nucleons averaged over the impact parameter was 
estimated to be $\langle A_{\text{part}} \rangle_{\text{b}} = 50$. 

\section{Data analysis}

Particles traversing the CDC detector activate sense wires along their flight 
path, leaving series of {\it hits}, which are collected into {\it tracks} by 
the off-line procedure. While hitting the MMRPC (PSB) detector, particles 
activate one or a few neighbouring strips, merged off-line into hits. 
Subsequently, tracks in the CDC and hits in the ToF detectors are matched. 
A collection of tracks is used to calculate the position $\vec v$ of the event
vertex. To reject the collisions occurred outside the target, a cut was 
applied on the component of the vertex position in the beam direction: 
$\left| v_{\text{z}} \right| < 15$~cm. 

In order to select a sample of good-quality tracks, a set of cuts was applied.
To suppress the contribution from the discontinuous tracks, the track was 
required to be constructed of at least 36 (32) hits for K$^-$ (K$^+$). The 
asymmetry in the number of hits is due to the non-radial profile of sense 
wires of the CDC, so the negative particles generate more hits on average than
the positive ones. As both charged kaons are produced in the target, two more 
conditions were imposed on transverse and longitudinal distance of closest 
approach between the track and the collision vertex: 
$\left| d_0 \right| < 1.5$~cm and $\left| z_0 \right| < 30$~cm, accordingly. 
Particles with the absolute values of charge higher than the elementary 
charge were rejected. Since some anisotropy was found in the azimuthal 
distribution of tracks in the backward zone of the CDC, the region of 
$\theta_{\text{lab}} > 90^\circ$ was excluded from the analysis. The 
requirement of CDC-ToF matching imposes a lower limit on the transverse 
momentum of a particle spiralling in the magnetic field: 
$p_{\text{t}} \geq 0.1$~GeV\slash c. To ensure the good matching between the 
track in the CDC device and the hit in a ToF detector, two additional 
conditions were imposed on the azimuthal and longitudinal distance between 
the extrapolation of the CDC track to the relevant ToF module, and the hit 
therein: 
$\left| \Delta \varphi \right| < 1.5^\circ$, 
$\left| \Delta z \right| < 30$~cm for the PSB, and
$\left| \Delta \varphi \right| < 0.6^\circ$, 
$\left| \Delta z \right| < 25$~cm for the MMRPC.

Charged particle identification based on information from the CDC and ToF 
detectors is performed by tracing the correlation between momentum and 
velocity of a particle, as shown in Fig.~\ref{fig:pvplot}. Events on this 
plane can be projected onto the mass parameter, using the relativistic 
formula $p = m\gamma v$. The identification capability of charged kaons, 
limited to modest momenta for the PSB, has been largely enhanced in the 
MMRPC, due to the excellent timing performance and better granularity of 
the latter~\cite{Kis11}. 

\subsection{Negative kaons}

For the analysis of K$^-$, basing on the observation of signal and background 
ratio the high $p_{\text{t}}$ limits were imposed of 0.57 GeV/$c$ (MMRPC) and
0.35 GeV/$c$ (PSB). To reconstruct the phase space population of K$^-$, the 
experimental mass distribution was analysed for every 
$p_{\text{t}}$-$y_{\text{lab}}$ cell, where $y_{\text{lab}}$ denotes rapidity 
in the laboratory frame, and the kaonic signal was separated from the 
background composed of $\pi^-$ mesons. A total of 9870 negative kaons were 
found in the CDC-ToF acceptance region. Influences of the choice of the 
binning of $p_{\text{t}}$, $y_{\text{lab}}$ and mass histograms, and of the 
choice of the minimum number of hits in a CDC track on the spectra presented
in this paper were included in the relevant systematic errors. 

\begin{figure}
 \includegraphics[width=8.6cm]{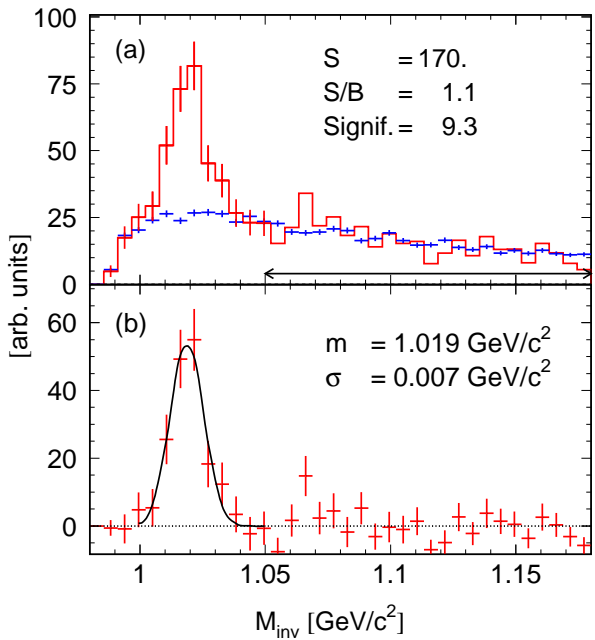}
 \caption{\label{fig:phiminv}(a) Invariant mass plot of (solid line) true, 
   and (scattered points) mixed K$^+$K$^-$ pairs. (b) $\phi$ meson signal 
   obtained after background subtraction.}
\end{figure}

\subsection{$\phi$ mesons}

For the reconstruction of the $\phi$ meson via the K$^+$K$^-$ decay, the 
maximum momentum of K$^+$ (K$^-$) was set to 1.2 (0.85) GeV/$c$ for the tracks
matched with the MMRPC, and 0.72 (0.60) GeV/$c$ for the PSB. To minimize the
side effects on the edges of ToF detectors, the observation region of the 
$\phi$ meson phase space was trimmed to 
$95^\circ < \theta_{\text{NN}} < 150^\circ$, where $\theta_{\text{NN}}$ is the
polar emission angle in the nucleon-nucleon centre of mass frame. 

The $\phi$ mesons were identified via the invariant mass analysis of K$^+$K$^-$
pairs (see Fig.~\ref{fig:phiminv}), a dominant $\phi$ meson decay channel. The
uncorrelated background was obtained with the mixed events method and 
normalized to the true pair distribution in the region 
$1.05 < M_{\text{inv}} < 1.18$~GeV/$c^2$. After subtraction of the background,
about 170 $\phi$ mesons were found under the peak. Within the range of $\pm$2
standard deviations of the fitted Gaussian distribution the signal to 
background ratio was found to be 1.1, and significance 9.3. The values of cut
parameters leading to the identification of $\phi$ mesons were further varied,
and the propagation of these changes on the results were included in their
systematic uncertainties. 

\section{Efficiency evaluation}

\begin{figure*}
 \subfigure{\includegraphics[scale=0.38]{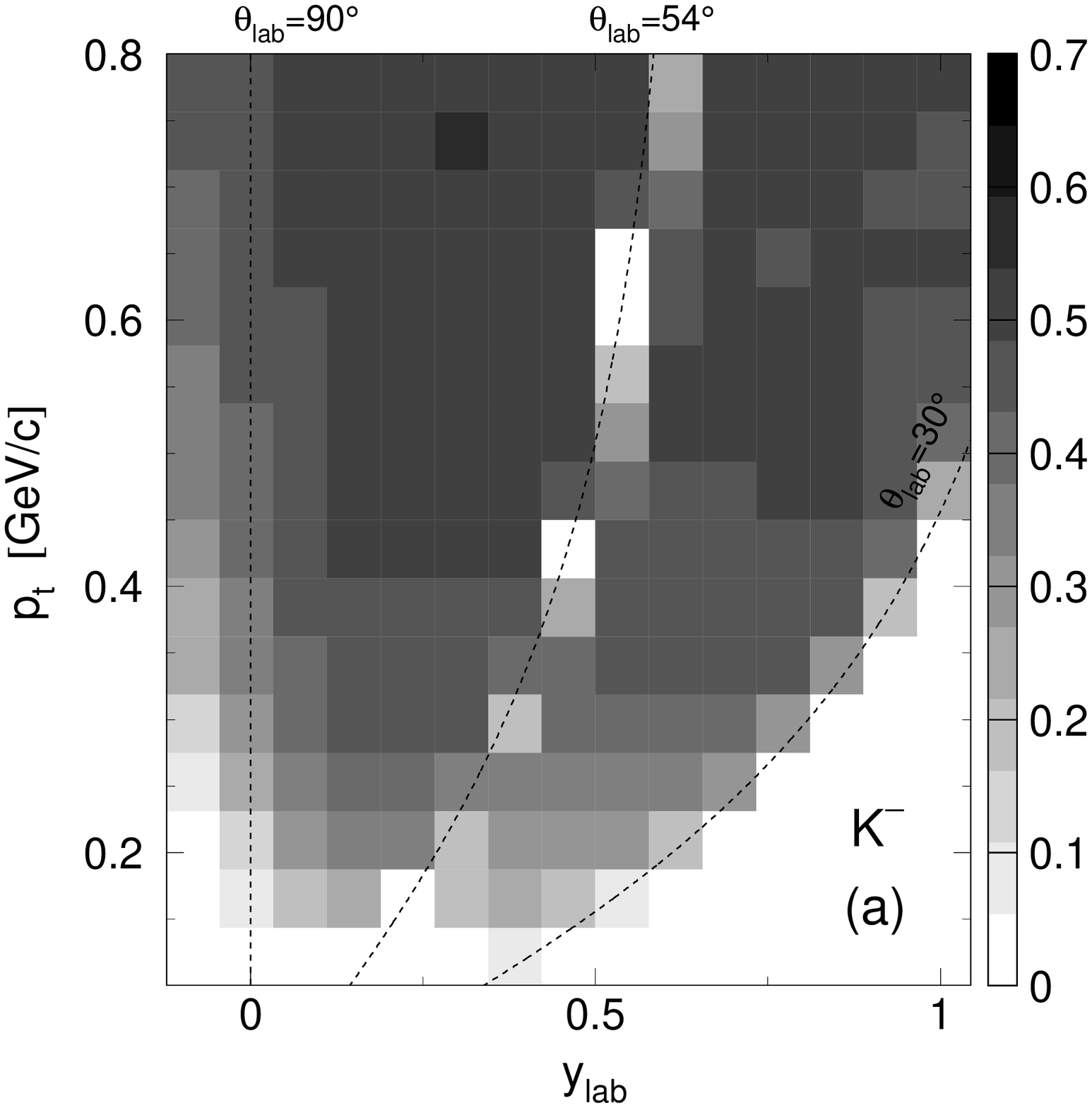}}\quad
 \subfigure{\includegraphics[scale=0.38]{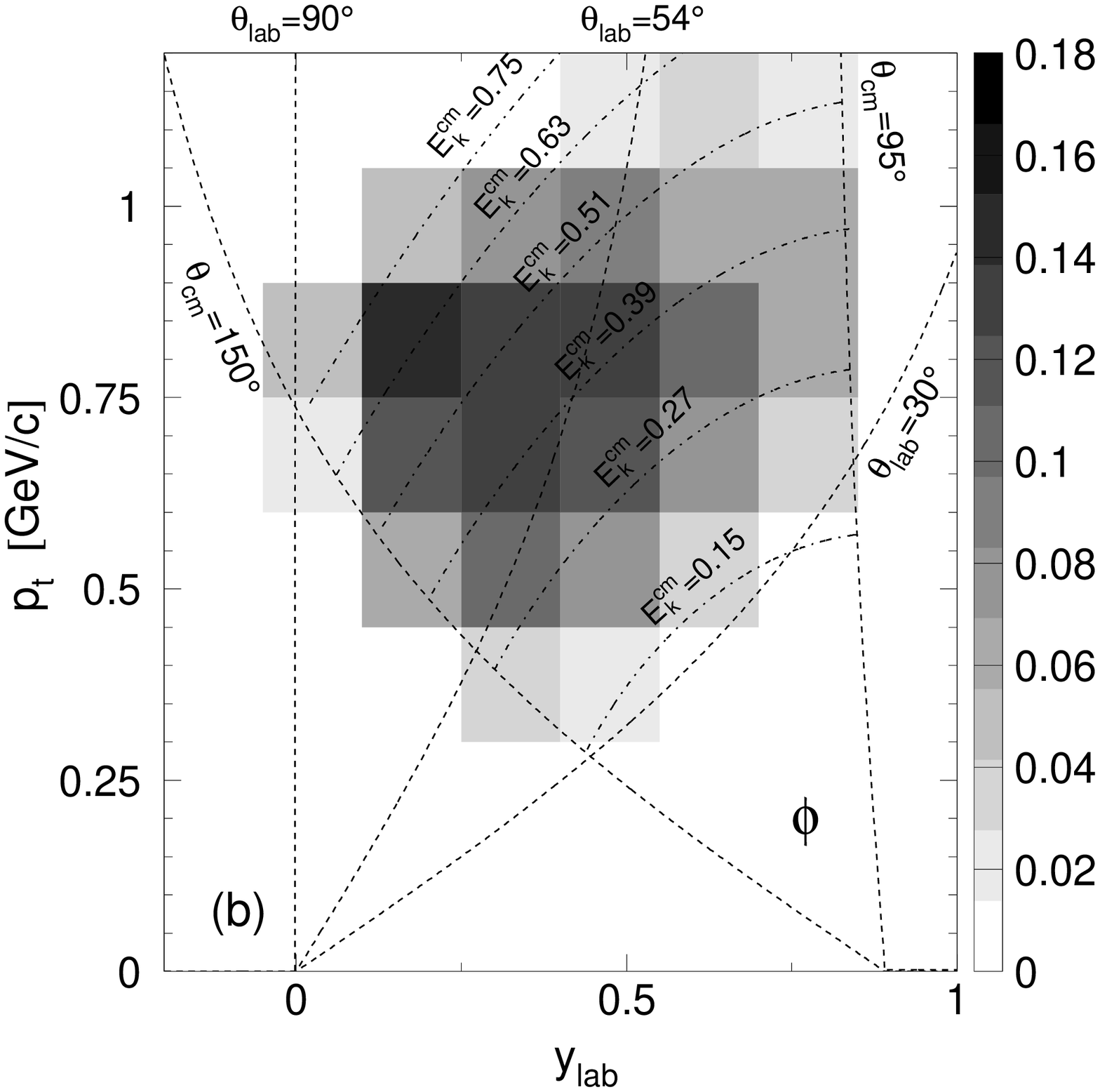}}
 \subfigure{\includegraphics[scale=0.38]{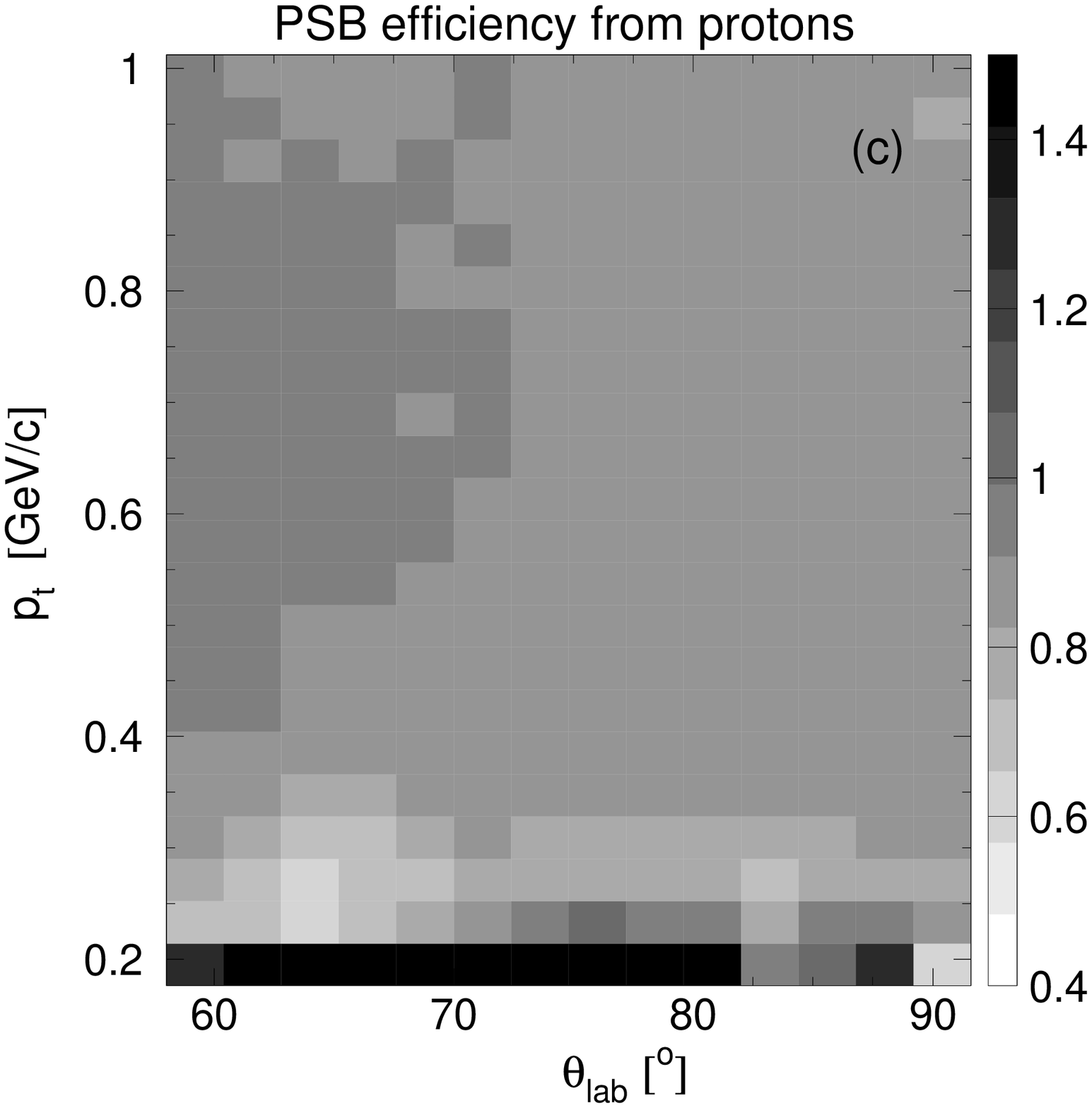}}\quad
 \subfigure{\includegraphics[scale=0.38]{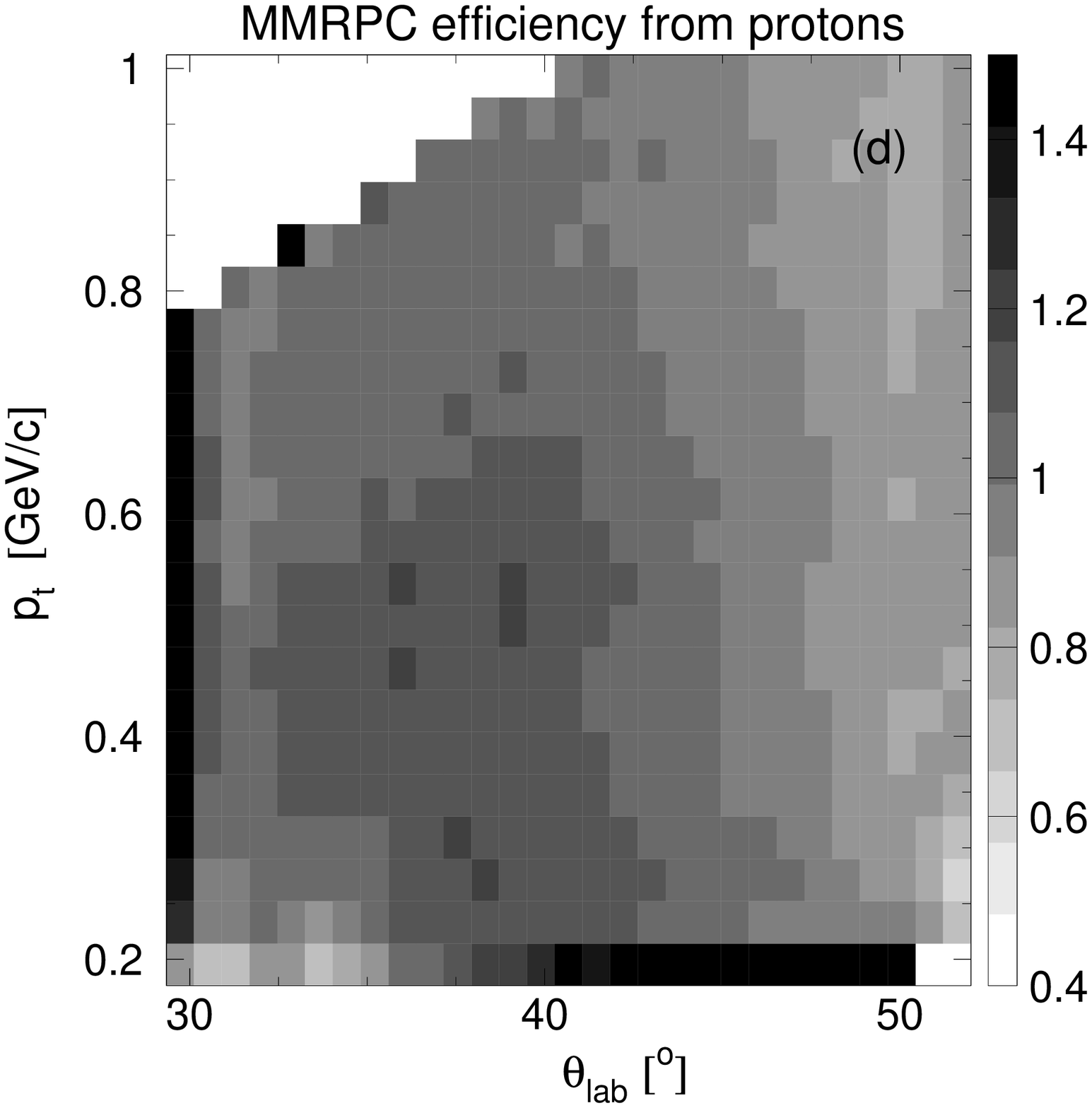}}
 \caption{\label{fig:efficiencies}Upper row: efficiency 
  distribution for (a) K$^-$, and (b) $\phi$ mesons (at 
  $T_{\text{s}} = 100$~MeV, $\alpha = 0$). Lower row: maps of internal 
  efficiency of (c) PSB, and (d) MMRPC detectors, obtained with protons. 
  See text for details.}
\end{figure*}

\begin{figure*}
 \subfigure{\includegraphics[scale=0.38]{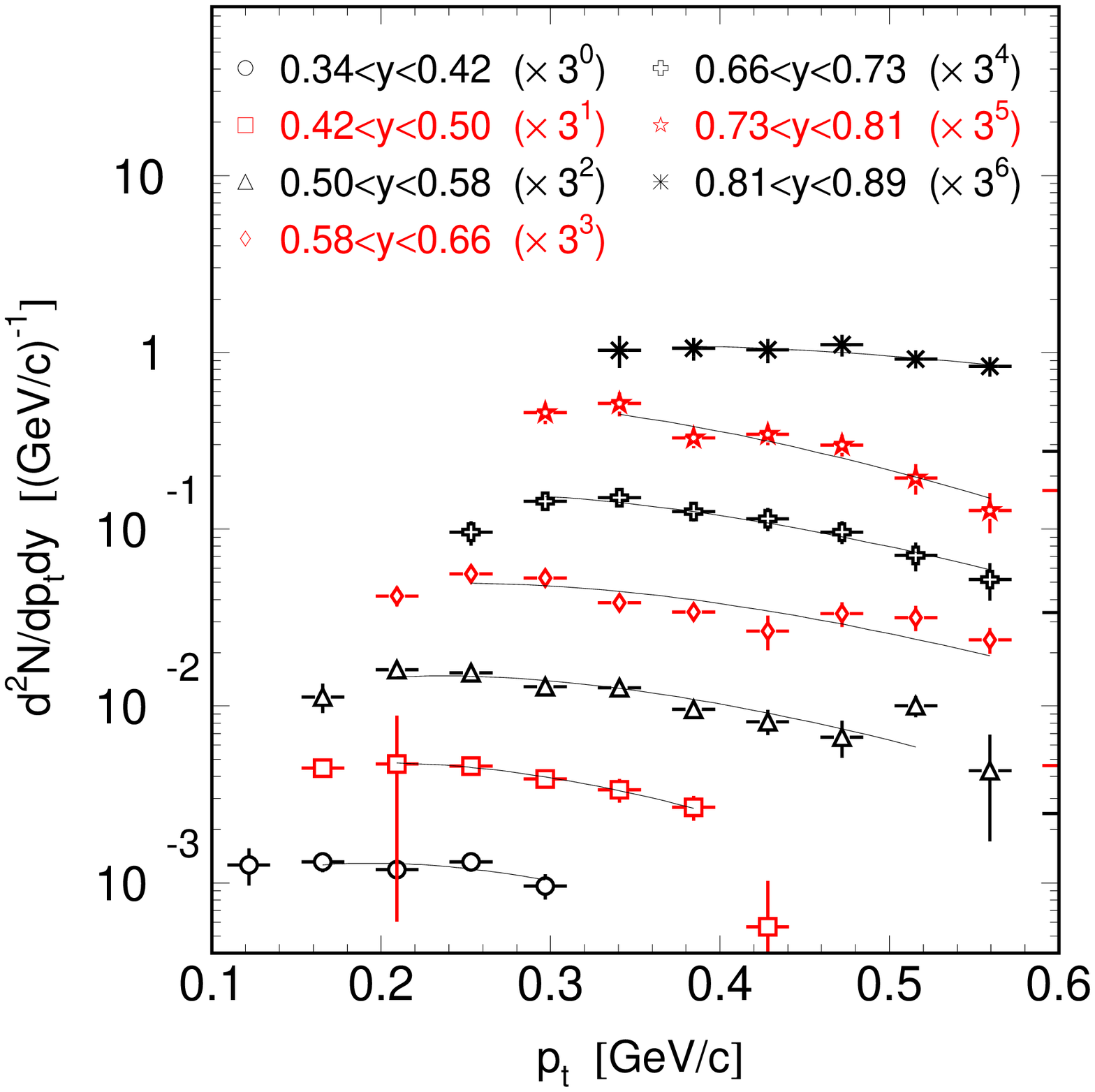}}\quad
 \subfigure{\includegraphics[scale=0.38]{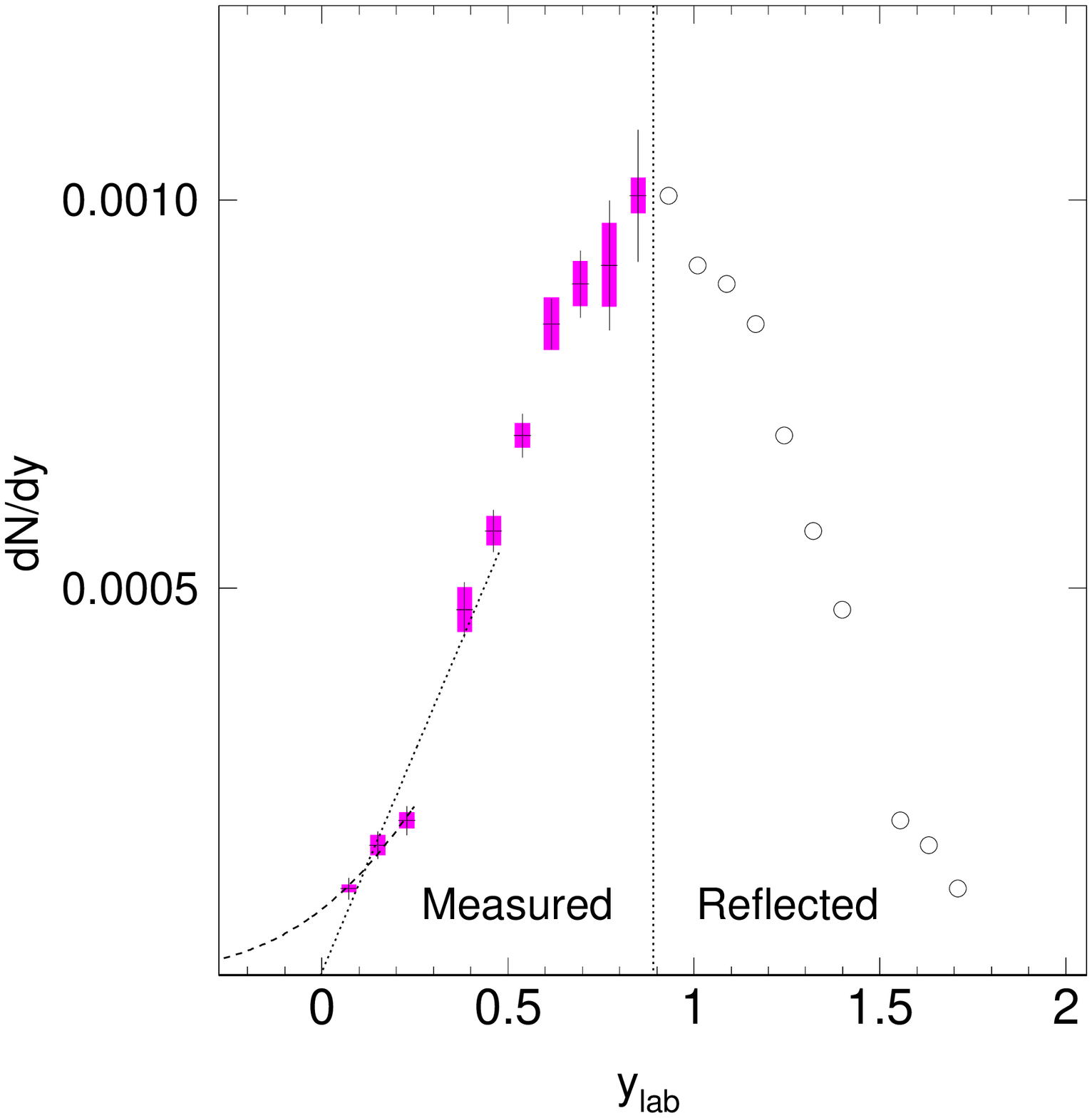}}
 \caption{\label{fig:kmphase}(Color online) (a) Transverse momentum 
 distributions of K$^-$ for the rapidity region $0.34 < y_{\text{lab}} < 0.89$.
 (b) Rapidity distribution of K$^-$. Data for $y_{\text{lab}}$ above 
 midrapidity (open circles) are the reflection of measured data points. 
 Boxes indicate systematic errors of points within the selected binning. 
 The dashed and dotted curves correspond to different extrapolation methods. 
 The vertical line indicates the centre of mass rapidity 
 $y^{\text{CM}}_{\text{NN}} = 0.89$. See text for details.}
\end{figure*}

\subsection{Standard efficiency correction}

\subsubsection{Negative kaons}

The GEANT~\cite{GEANT} package for detector simulation was employed to obtain
the efficiency correction. Negative kaons were generated according to the
homogeneous \mbox{$p_{\text{t}}$-$y_{\text{lab}}$} distribution, and embedded
in the events of products of the Ni+Ni collisions at the beam energy of 1.91A 
GeV, simulated by the IQMD code~\cite{IQMD}. For both K$^-$ and $\phi$ mesons, 
the simulated events were treated using the same off-line analysis package, as 
for the experimental data. The efficiency distribution of K$^-$ for the regions 
of phase space covered by MMRPC and PSB detectors is shown in 
Fig.~\ref{fig:efficiencies}a. For the momenta higher than 0.5~GeV/$c$ it 
reaches about 50\%. A drop of efficiency at lower $p_{\text{t}}$ is caused 
by the higher probability for the kaon to decay on its path to either of two 
ToF detectors. 

\subsubsection{$\phi$ mesons}

For the efficiency evaluation, the mass of the $\phi$ mesons was sampled from 
the Breit-Wigner distribution, and the phase space was populated by pulling from 
the Boltzmann distribution scaled by the anisotropy term
\begin{equation}
 \label{EQ:phigeantmodel}
 \frac{d^2N}{dEd\vartheta} \sim ~pE~ \exp (-E\slash T_{\text{s}})  \cdot \left(1 + \alpha \cos^2 \vartheta\right) ,
\end{equation}
\noindent where $T_{\text{s}}$ is the temperature of the source, and $\alpha$ 
is the anisotropy parameter. Sampled mesons were subsequently boosted to the 
laboratory frame. The parameters in the Eq.~\ref{EQ:phigeantmodel} were varied 
in the range of $T_{\text{s}} \in$~\mbox{[80, 130]~MeV}, 
$\alpha \in$~\mbox{[0, 0.6]}. Differences of the obtained efficiency 
corrections due to variations of these parameters were further included in 
the systematic uncertainties of the investigated physics variables, as 
discussed below. $\phi$ mesons were subsequently embedded in the Ni+Ni 
collisions, simulated with the IQMD code. Fig.~\ref{fig:efficiencies}b shows 
the phase space efficiency distribution for $T_{\text{s}} = 100$~MeV, and 
$\alpha = 0$. As the $\phi$ mesons have been analysed both inclusively and 
by studying the kinetic energy distribution, the appropriate efficiencies 
were obtained for either case separately. 

\subsection{Internal efficiency of ToF detectors}

Following the reported inhomogeneity in the longitudinal position response of
the MMRPC detector, shown in Fig.~12 of Ref.~\cite{Kis11}, the phenomenological
study of the internal efficiency of this detector was performed. While the 
effects of geometry and matching are the regular part of the GEANT-based 
efficiency determination procedure, possible internal MMRPC inefficiencies 
were not included so far. This additional efficiency factor was pursued by 
constructing first the ratio of CDC tracks with associated hit in a ToF 
(MMRPC, PSB) detector to all the reconstructed CDC tracks. This ratio was 
obtained independently for the experimental and simulated data. Next, both 
results were divided, to yield the internal efficiency factor $f$, according to:

\begin{equation}
f^{ToF} \left( \vartheta, p_{\text{t}} \right) 
 = \frac{N^{\text{ToF}}_{\text{exp}}}{N^{\text{CDC}}_{\text{exp}}}
   \slash
   \frac{N^{\text{ToF}}_{\text{sim}}}{N^{\text{CDC}}_{\text{sim}}}
\end{equation}

\noindent where ToF~$\in$~(MMRPC, PSB). The resulting 
f($\vartheta$,~$p_{\text{t}}$) maps for both PSB and MMRPC detectors are shown
in Fig.~\ref{fig:efficiencies}. Particularly noticeable is the heap structure
around the centre of length of the MMRPC detector ($\vartheta \approx 38^\circ$),
reported also in~\cite{Kis11}. A correction of the distributions of K mesons 
was done on an event-by-event basis by weighting every kaon event with the 
factor \mbox{$1 \slash f(\vartheta, p_{\text{t}})$}. In order to minimize the 
possible sensitivity of $f$ to particle's charge, corrections for negatively 
charged kaons were performed using the map obtained from $\pi^-$ mesons. 
Differences between maps obtained from protons, and $\pi^-$ were found to be 
small (in the order of 5-7\%) and were manifested mainly in the global 
normalization. 

\section{Results}

\subsection{Negative kaons}

The transverse momentum spectra for consecutive slices of rapidity within the 
$0.34 < y_{\text{lab}} < 0.89$ range, covered by the MMRPC, are shown in 
Fig.~\ref{fig:kmphase}a. They were fitted according to the Boltzmann-like 
function:

\begin{equation}
 \label{EQ:ptfit}
 \frac{d^2 N}{dp_{\text{t}} dy_{\text{NN}}} = N ~p_{\text{t}} 
              E ~\exp (-m_{\text{T}} \slash T_{\text{B}} )
\end{equation}

\noindent where $m_{\text{T}} = \sqrt{p_{\text{T}}^2 + m^2}$ is the transverse
mass, and for every slice $y_{\text{NN}}$ denotes the average value of rapidity
in the NN frame, $N$ is the normalization factor, and $T_{\text{B}}$ is the 
inverse slope (also called the apparent temperature). The $p_{\text{t}}$ 
spectra measured in the PSB region ($0 < y_{\text{lab}} < 0.3$) had too poor 
statistics for the two-parameter fitting. Therefore, the normalization 
parameters were extracted only, by fitting the formula \ref{EQ:ptfit} with one 
of two fixed values of $T_{\text{B}}$: 45 and 60~MeV (differences in the 
results were accounted for in the evaluation of the systematic errors). The 
points of the rapidity distribution were obtained by an analytic integration
of the formula \ref{EQ:ptfit} from 0 to $\infty$ for every slice,

\begin{equation}
 \frac{dN}{dy_{\text{NN}}} = N \cdot \cosh (y_{\text{NN}}) \cdot T_{\text{B}}^3 
                 \left( \frac{m^2}{T_{\text{B}}^2} + 2\frac{m}{T_{\text{B}}} + 2 \right)
                 \exp \left( -\frac{m}{T_{\text{B}}} \right)   \  , 
\end{equation}

\noindent where $m$ is the particle's mass, and substituting the parameters 
obtained in the fit above. The obtained distribution is shown in 
Fig.~\ref{fig:kmphase}b. Note, that the N and T$_{\text{B}}$ fit parameters are 
typically characterized by the strong anticorrelation term in the covariance 
matrix, which was included in the error evaluation of the rapidity 
distribution.The magenta boxes on the abovementioned plot correspond to the 
systematic errors arising from variations of all the applied cuts and binnings,
except for the contribution from the choice of binning of the rapidity axis. 
Benefitting from the symmetry of the colliding system, the measured data points
were reflected with respect to the midrapidity 
(y$^{\text{CM}}_{\text{NN}} = 0.89$), which allowed for a wide coverage of 
the K$^-$ rapidity spectrum. 

In order to obtain the K$^-$ yield, the tails of the rapidity distribution 
were extrapolated. To assess the systematic error of this procedure, the data 
was fitted using the Gaussian and linear functions, each one in two ranges: 
$y_{\text{lab}} < 0.3$, and 0.5. The variant resulting in the largest 
(smallest) reconstructed yield is shown as dashed (dotted) line in 
Fig.~\ref{fig:kmphase}b. The total yield of K$^-$ was found to be:

\begin{equation}
 \rm{P(K^-) = (9.84 ~\pm~ 0.21~(stat) ~^{+0.63}_{-0.57}~(syst)) \times 10^{-4}}
\end{equation}

\noindent per triggered event. Note, that the relatively wide coverage of 
phase space allowed to minimize the assumptions on the overall shape profile 
of the rapidity distribution, and in consequence on the total yield of K$^-$. 

\subsection{$\phi$ mesons}

The limitation of the sample of $\phi$ mesons to about 170 events does not 
permit for a full-fledged analysis of its phase space, and thus for the 
reconstruction of the yield in a model-independent fashion. As a first step 
the yield was reconstructed directly by dividing the total number of events 
within the phase space region reported above by the inclusive efficiency for 
this region. However, the result revealed a clear correlation with the input 
temperature $T_{\text{s}}$ of the source simulated in the efficiency evaluation
procedure (cf. Eq.~\ref{EQ:phigeantmodel}). It ranged between 
$3.2 \times 10^{-4}$ for $T_{\text{s}} = 80$~MeV, and $6.2 \times 10^{-4}$ for
$T_{\text{s}} = 130$~MeV, not accounting for any statistical or systematic 
errors. This uncertainty was a motivation to study the kinetic energy spectrum
of $\phi$ mesons, in hope that, apart from carrying the kinematical information
itself, it may provide constraints on the temperature, and thus limit the 
uncertainty of the total yield. The reconstructed kinetic energy spectrum, 
was fitted with the Boltzmann-like function of the form 

\begin{figure}[t]
 \includegraphics[width=8.6cm]{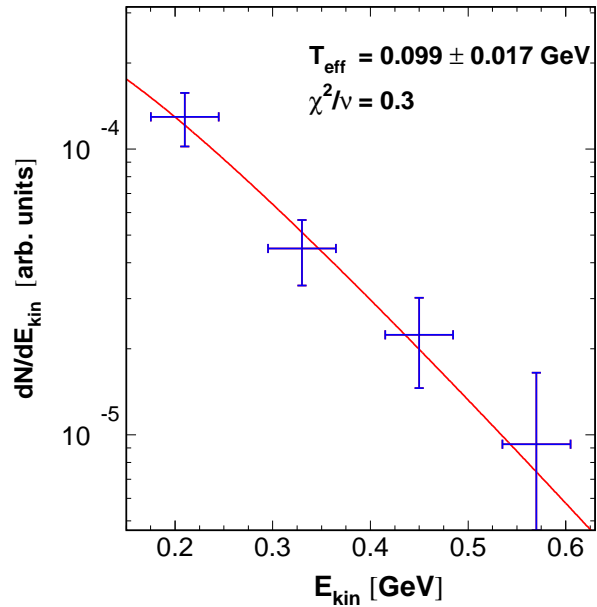}
 \caption{\label{fig:phiekin}(Color online) Kinetic energy distribution of 
$\phi$ mesons fitted with Boltzmann-like function, assuming the following 
input parameters to the efficiency calculation: $T_{\text{s}} = 100$~MeV, 
$\alpha = 0$ (see text for details).}
\end{figure}

\begin{equation}
 \label{EQ:ekfit}
 \frac{dN}{dE_{\text{k}}} = N ~pE~ \exp (-E\slash T_{\text{eff}}) ,
\end{equation}

\noindent where $N$ is the normalization parameter, total energy 
$E = E_{\text{k}} + m$, momentum $p = \sqrt{E^2 - m^2}$, and $T_{\text{eff}}$ 
is called the inverse slope. An exemplary case of fitting the kinetic energy 
spectrum at $T_{\text{s}} = 100$~MeV, $\alpha = 0$, and events grouped in four 
bins, is shown in Fig.~\ref{fig:phiekin}. The extracted value of 
$T_{\text{eff}}$ was found to depend a little on a choice of binning of the 
$E_{\text{k}}$ spectrum, $T_{\text{s}}$ and $\alpha$ efficiency input 
parameters, particle identification cuts, and fit range imposed. However, only
the fits obtained with the input $T_{\text{s}}$ around the middle of the 
probed range of $~\mbox{[80, 130]~MeV}$ resulted in values of $T_{\text{eff}}$
fit parameter consistent with $T_{\text{s}}$. Following this finding, a 
self-consistency condition was imposed: 
$\left| T_{\text{s}} - T_{\text{eff}} \right| < 15$~MeV, where 15~MeV is the 
typical statistical error of the fitted $T_{\text{eff}}$. Note, that this 
condition not only narrows down the ranges of input $T_{\text{s}}$ parameter, 
and the systematic error of $T_{\text{eff}}$, but the selection of the 
$T_{\text{s}}$ region also limits the systematic error of the total 
$\phi$-meson yield discussed above. Within the imposed condition, the inverse 
slope of the kinetic energy distribution of $\phi$ mesons was found to be 
$T_{\text{eff}} = 105 \pm 18 \text{(stat)} ^{+19}_{-13} \text{(syst)}$. The 
total yield was extracted by extrapolation of the data points with the fitted
curve (Eq.~\ref{EQ:ekfit}), and found to be 

\begin{equation}
 \rm{P(\phi) = (4.4 ~\pm~ 0.7 ~\text{(stat)} ~^{+1.7}_{-1.4} ~(syst)) 
     \times 10^{-4}}
\end{equation}

\noindent per triggered event. We found, that the impact of non-zero anisotropy 
parameter $\alpha$ on the final results is on the level of 5\% of this value.

\subsection{Influence of $\phi$ production on K$^-$ yields}

The nearly 50\% branching ratio of the $\phi \rightarrow \text{K}^+ \text{K}^-$ 
decay channel, and the comparable yields of $\phi$ and K$^-$ mesons suggest that 
their emission should be considerably correlated. However, while direct K$^-$ 
are emitted from the hot collision zone, most decay products of $\phi$ mesons 
are created outside this region. Therefore, one may expect that negative kaons 
emitted from both these sources could have different kinematical 
characteristics. Hence, the question arises, what fraction of the observed 
K$^-$ originate from the decays of $\phi$ mesons. For this analysis we find the
following ratio of yields: 

\begin{equation}
 \rm{\frac{P(\phi)}{P(K^-)} = 0.44 \pm 0.07 (stat) ^{+0.18}_{-0.12} (syst) 
     \quad ,} 
\end{equation}

\noindent where the systematic errors were obtained by combining all the 
variations of both components due to their own systematic uncertainties, and 
rejecting 5\% of values on the tails of the resulting distribution. Taking into
account that 48.9\% $\phi$ mesons decay into K$^+$K$^-$ pairs, it translates 
into $22 \pm 3 ^{+9}_{-6} ~\%$ of observed K$^-$ originating from decays of 
$\phi$ mesons. It is worthwhile noting that the same values within errors have
been obtained for the Ar+KCl system colliding at even more subthreshold beam
energy of 1.756A GeV~\cite{Agak09}. Also, for the central Al+Al collisions at
1.9A GeV, a similar value of \mbox{$\text{P}(\phi)\slash \text{P}(\text{K}^-)$} 
$= 0.30 \pm 0.08 \text{(stat)} ^{+0.04}_{-0.06} \text{(syst)}$ 
was found~\cite{Gasi10,GasiTh}. However, for the elementary pp collisions at 
2.7 GeV, this ratio was found to be slightly above 1~\cite{Maed08}. One 
possible explanation of this discrepancy arises in the context of the 
statistical model where the volume of open strangeness production is assumed 
to be limited and parametrized by the canonical radius $R_{\text{C}}$. 
According to the calculations reported in~\cite{Agak09}, the $\phi$\slash 
K$^-$ ratio measured for Ar+KCl is consistent with $R_{\text{C}}$ between 2.2 
and 3.2 fm, while that for pp is reproduced for $R_{\text{C}} \approx 1.2$~fm,
although the latter prediction disagrees with the experimental data at higher
beam energies. Unfortunately, performing the statistical model fit to the data
was not possible for the present experiment, due to a meager amount of 
reconstructed particle yields available at centralities corresponding to 
$\langle A_{\text{part}} \rangle_{\text{b}} = 50$. Much wider array of particle
yields is available for the central collisions of the studied system (c.f. 
Fig.~4b in~\cite{Pias09}). A measurement of \mbox{$\phi$\slash K$^-$} ratio 
for this centrality range would give an opportunity to extract the canonical
radius parameter for the Ni+Ni system and compare with that for the Ar+KCl 
collisions.

\begin{figure}[b]
 \includegraphics[width=8.6cm]{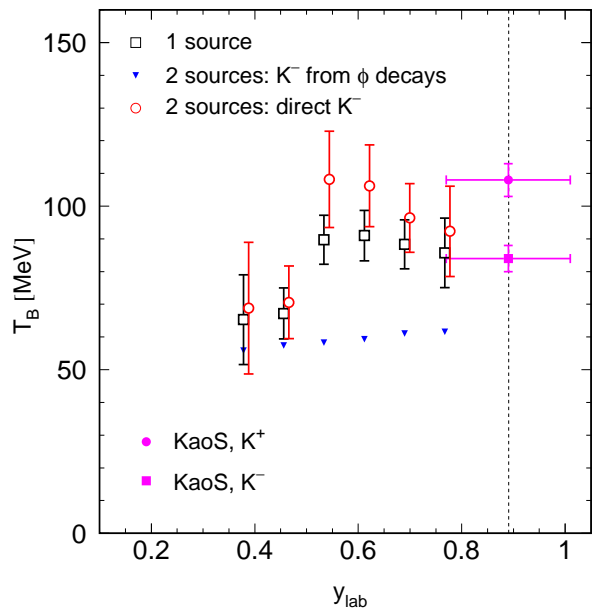}
 \caption{\label{fig:kaon2src}(Color online) Rapidity distribution of 
 $T_{\text{B}}$ (apparent temperatures) for negative kaons within the 
 one-source hypothesis (open squares), and two-source approach (K$^-$ from 
 $\phi$ decays - full triangles, direct kaons - open circles). Inverse slope 
 of K$^+$ (K$^-$) obtained by the KaoS Collaboration is marked by full circle 
 (square). See text for details. }
\end{figure}

In order to estimate the influence of $\phi$ mesons on the kinematic properties
of the observed K$^-$, PLUTO code~\cite{PLUTO} was first employed to generate
the thermal, and isotropic $\phi$ mesons with temperature of 105~MeV, which
subsequently decayed, giving rise to K$^-$ production. The inverse slopes of
the $p_{\text{t}}$ spectra of such kaons, shown in Fig.~\ref{fig:kaon2src} in
full triangles, were found to be clearly lower than those of experimentally
measured K$^-$, depicted with open squares. Following this finding, the 
emission of negative kaons was assumed to arise from two sources: the collision
zone emitting kaons directly, and $\phi$ mesons decaying in vacuum. Their 
contributions were weighted according to the experimentally found 
\mbox{$\phi$\slash K$^-$} ratio, and the phase space distributions of both were
assumed to be Boltzmann-like. For every slice of rapidity, the temperature
parameter (inverse slope) of kaons from the $\phi$-meson decays was fixed at a
value obtained from the simulation described above, while the slope for the
direct kaons was extracted by fitting the two-source model to the experimental
data, giving the results shown in open circles in Fig.~\ref{fig:kaon2src}. 
Keeping in mind the limited statistics, and simplicity of the model, the 
results suggest that removing 22\% contribution from $\phi$ mesons could 
systematically raise the inverse slope of K$^-$ by up to about 15~MeV. 

In contrast, as the K$^-$\slash K$^+$ yield ratio is about 3\%~\cite{Fors07},
the influence of $\phi$ mesons on positively charged kaons is negligible. The 
inverse slope parameters of the kinetic energy distributions of K$^+$ and 
K$^-$, obtained by the KaoS Collaboration at midrapidity for the same 
colliding system, and similar $\langle A_{\text{part}} \rangle_{\text{b}}$, 
were found to exhibit a gap of about 25~MeV. Our finding suggests, that the 
contribution from $\phi$ mesons to K$^-$ emission significantly ,,cools down"
the overall spectrum of negative kaons. Thus, this effect may account for a
sizeable share of this gap, possibly competing with modifications of kaonic
properties in the nuclear medium. 

\section{Summary}

Production of $\phi$ and K$^-$ mesons was investigated in Ni+Ni collisions
at the beam kinetic energy of 1.91A GeV. The trigger selected a sample of
central and semi-central collisions amounting to 51\% of the geometrical 
cross section. The $p_{\text{t}}$ and $y_{\text{lab}}$ distribution of 
K$^-$ were analysed in a wide region of phase space. The total yield of
K$^-$ was found to be $(9.84 \pm 0.21~(\text{stat}) ^{+0.63}_{-0.57}~
(\text{syst})) \times 10^{-4}$ per triggered event. About 170 $\phi$ mesons
were reconstructed. The inverse slope of kinetic energy distribution was
found to be $T = 105 \pm 18 (\text{stat}) ^{+19}_{-13} (\text{syst})$~MeV,
and the total yield $(4.4 \pm 0.7 (\text{stat}) ^{+1.7}_{-1.4} 
~(\text{syst})) \times 10^{-4}$ per triggered event. 

The found $\phi$\slash K$^-$ ratio of $0.44 \pm 0.07 (\text{stat}) 
^{+0.18}_{-0.12} (\text{syst})$ means that $22 \pm 3 \, ^{+9}_{-6} ~\%$ of 
K$^-$ originate from decays of $\phi$ mesons, occurring mostly in vacuum.
The influence of this additional source of negative kaons on the transverse
momentum spectra was studied within a two-sources approach, where the 
contribution from $\phi$-meson decays was modelled by an isotropic 
Boltzmann-like distribution. The inverse slopes of K$^-$ produced directly
in the collision zone seem to be up to about 15 MeV higher than the values
extracted within the one-source hypothesis. This effect, compared to the 
25 MeV gap between the inverse slopes of K$^+$ and K$^-$, signals that a 
considerable share of the gap could be explained by feeding of negative 
kaons by the $\phi$ meson decays. Thus, it seems crucial to account for 
the contribution of $\phi$-originating negative kaons in the studies of 
the in-medium modifications of these particles via comparisons of 
K$^-$\slash K$^+$ or flow (e.g.~$v_{1,2}$) distributions to the predictions
of the transport codes. 

\begin{acknowledgments}
This work was supported by the German BMBF Contract No. 05P12VHFC7, the Korea
Science and Engineering Foundation (KOSEF) under Grant No. 
F01-2006-000-10035-0, by the German BMBF Contract No. 05P12RFFCQ, by the 
Polish Ministry of Science and Higher Education (DFG/34/2007), the agreement
between GSI and IN2P3/CEA, the HIC for FAIR, the Hungarian OTKA Grant No. 
71989, by NSFC (Project No. 11079025), by DAAD (PPP D/03/44611), by DFG 
(Projekt 446-KOR-113/76/04) and by the EU, 7th Framework Program, Integrated
Infrastructure: Strongly Interacting Matter (Hadron Physics), Contract No. 
RII3-CT-2004-506078.
\end{acknowledgments}

\end{document}